\documentclass[iop,apj,tighten,numberedappendix]{emulateapj}

\usepackage[breaklinks,colorlinks,urlcolor=blue,citecolor=blue,linkcolor=blue]{hyperref}

\usepackage{amsmath,amssymb}
\usepackage{cleveref}
\usepackage{graphicx}
\usepackage{bm}
\usepackage[toc,title,page]{appendix}
\usepackage{tocbibind}
\usepackage{color}
\newcommand{\Msun}{{\rm M}_\odot}

\usepackage{xcolor}

\shorttitle{Collisions between stars and AGN disks}
\shortauthors{Tagawa and Haiman}

\begin{document}
\title{
Flares from stars crossing active galactic nuclei disks on low-inclination orbits}

\author{
Hiromichi Tagawa\altaffilmark{1,2,3}, 
Zolt\'an Haiman\altaffilmark{1,4}
}
\affil{
\altaffilmark{1}Department of Astronomy, Columbia University, 550 W. 120th St., New York, NY, 10027, USA\\
\altaffilmark{2}Astronomical Institute, Graduate School of Science, Tohoku University, Aoba, Sendai 980-8578, Japan. \\
\altaffilmark{3}
National Astronomical Observatory of Japan, National Institutes of Natural Sciences, 2-21-1 Osawa, Mitaka, Tokyo 181-8588, Japan. \\
\altaffilmark{4}Department of Physics, Columbia University, 550 W. 120th St., New York, NY, 10027, USA\\
}
\email{E-mail: htagawa@astr.tohoku.ac.jp}

\begin{abstract} 
The origin of the recently discovered new class of transients, X-ray quasi-periodic eruptions (QPEs), remains a puzzle. 
Due to their
periodicity and association with active galactic nuclei (AGN), 
it is natural to relate these eruptions 
to stars or compact objects in tight orbits around supermassive black holes (SMBHs). 
In this paper, we predict the properties of emission from bow shocks 
produced by stars crossing AGN disks,
and compare them to the observed properties of QPEs. 
We find that when a star's orbit is retrograde and has
a low inclination 
($\lesssim 40^\circ$) 
with respect to the AGN disk 
and the star is massive ($\gtrsim 10~\Msun$), 
the breakout emission from the bow shock can explain the observed duration ($\sim$ hours) and X-ray luminosity ($\sim$few$\times10^{42}~{\rm erg~s^{-1}}$) of QPEs.
This model can further explain various observed features of QPEs, 
such as their complex luminosity evolution, 
the gradual decline of luminosity 
of the flares over several years, 
the evolution of the hardness ratio, 
the modulation 
of the luminosity during quiescent phases, 
and the preference of the central SMBHs to have low masses. 
\end{abstract}

\section{Introduction}

Recently a new class of transients, X-ray quasi-periodic eruptions (QPEs), have been discovered. 
QPEs are characterized by periodic flares with the X-ray luminosity of $L_{\rm X}\sim 10^{42}$--$10^{43}~{\rm erg/s}$ and thermal-like spectra with the temperature of $\sim 100~{\rm eV}$. The duration of each flare is about an hour followed by a quiescent phase with the luminosity of $L_{\rm X}\sim 10^{41}~{\rm erg/s}$ and the duration of about ten hours. 
The period is quite regular and the cycle 
repeats 
for at least several years 
although some modulations and evolution are also reported \citep{Miniutti2022,Arcodia2022}. 
So far, five QPE events have been reported, whose emission is coming from the central regions of low-mass galaxies harbouring less massive supermassive black holes (SMBHs) with masses of $M_{\rm SMBH}\sim 10^5$--$10^6~\Msun$. 
All of the host galaxies have been classified as active by re-observing them using high-resolution optical spectroscopy \citep{Wevers2022}, although no broad emission lines, common in active galactic nuclei (AGNs), have been detected. 

The first QPE source, GSN069, was reported from a Seyfert-2 galaxy \citep{Miniutti2019}. 
Thereafter, the second and fifth QPE source, 
RX~J1301.9+2747 \citep{Giustini2020} and XMMSL1~J024916.6-041244 \citep{Chakraborty2021}, 
were discovered by XMM-Newton 
\citep{Jansen2001}, 
and the third and fourth, 
eRO-QPE1 and eRO-QPE2 \citep{Arcodia2021}, by the wide-field X-ray telescope eROSITA \citep{Predehl2021}. 
It is remarkable that the QPEs in GSN069 and XMMSL1~J024916.6-041244 followed flares interpreted as tidal disruption events \citep{Miniutti2019,Chakraborty2021,Miniutti2022}. 
Among the five QPEs, eRO-QPE1 exhibits a complex evolution of its light curve \citep{Arcodia2022}. 

So far, many theoretical models have been proposed. 
One possibility is disk instability. 
Various types of instabilities have been considered, 
such as those related to 
viscous torques \citep{Pan2022,Pan2023}, 
magnetic pressure \citep{Sniegowska2022,Kaur2022}, and spin-disk misalignment \citep{Raj2021}. However, 
it is not clear why these peculiar AGNs exhibit instability while other AGNs and X-ray binaries do not (but see \citealt{Kaur2022}).

Another class of models involves mass transfer \citep{Zalamea2010,Krolik2022}. 
To produce the high luminosity of QPEs in these models, 
high-density objects such as white dwarfs or helium stars are required \citep{King2020,Zhao2022}. 
Additionally, in order for the lost mass to be accreted onto the central SMBH during the active phases of about an hour, the eccentricity needs to be as high as $\sim 0.99$ for white dwarfs \citep{King2020,King2022,Wang2022}. 
However these configurations are estimated to be too rare to explain the observed rate of QPEs \citep{Metzger2022,Lu2022}. 
\citet{Metzger2022} proposed that mass transfer between stars counter-rotating with respect to each other around an SMBH can also explain the observed properties. 
\citet{Lu2022} suggested that even without high density objects and highly eccentric orbits, thermal emission from shocks caused by the collision 
between the envelope of a star 
and 
the disk formed by the mass lost during preceding orbits from the star can account for 
the observed luminosity. They also explain the pathway to realize this configuration  (see also \citealt{Linial2022_QPE}). However, the duration of emission is typically uncomfortably long 
due to the slow diffusion of photons in this model. 
Although advection 
has been suggested to reduce the timescale, it is not obvious whether photons can efficiently escape from an optically thick medium without reducing the luminosity and/or temperature.

A third possibility is emission from shocks emerging from collisions between an AGN disk and a star \citep{Xian2021,Sukova2021}. suggested that the possible modulation of the recurrence time between active phases can be explained by precession of stellar motion due to general relativistic effects in the deep potential well near the central SMBH. 
\citet{Xian2021} also estimated the luminosity of the flares using formulae in the optically thin limit provided in \citet{Nayakshin2004}. 
However, in optically thin cases, the luminosity can not be as high as the observed value; 
\citet{Nayakshin2004} finds an upper limit of around $\sim 10^{39}~{\rm erg/s}$ 
(see their Eq.~37), 
adopting the stellar position of $\sim 100~r_{\rm g}$ with the AGN disk temperature of $\sim 10^6~{\rm K}$ to
match the duration and spectrum of the observed QPEs. 
(Here $r_{\rm g}=GM_{\rm SMBH}/c^2$ is the gravitational radius, $G$ is the gravitational constant, and $c$ is the speed of light.) 
\citet{Dai2010} predicted the properties of emission from stellar collisions focusing especially on the timing of the flares. 
On the other hand, 
previous studies 
do not take into account emission from shocks breaking out of the AGN disk, which has properties significantly different from their estimates. 
Additionally, 
stars on low-inclination orbits with respect to the AGN disk enhance the total heating rate of gas, 
which helps explain the high observed luminosity of QPEs.

In this paper, we investigate properties of breakout emission from shocks produced by collisions between a star and an AGN disk, by constructing a semi-analytical model. Then we discuss whether the properties of QPEs can be explained by this collision and shock breakout scenario. 

\begin{figure}
\begin{center}
\includegraphics[width=85mm]{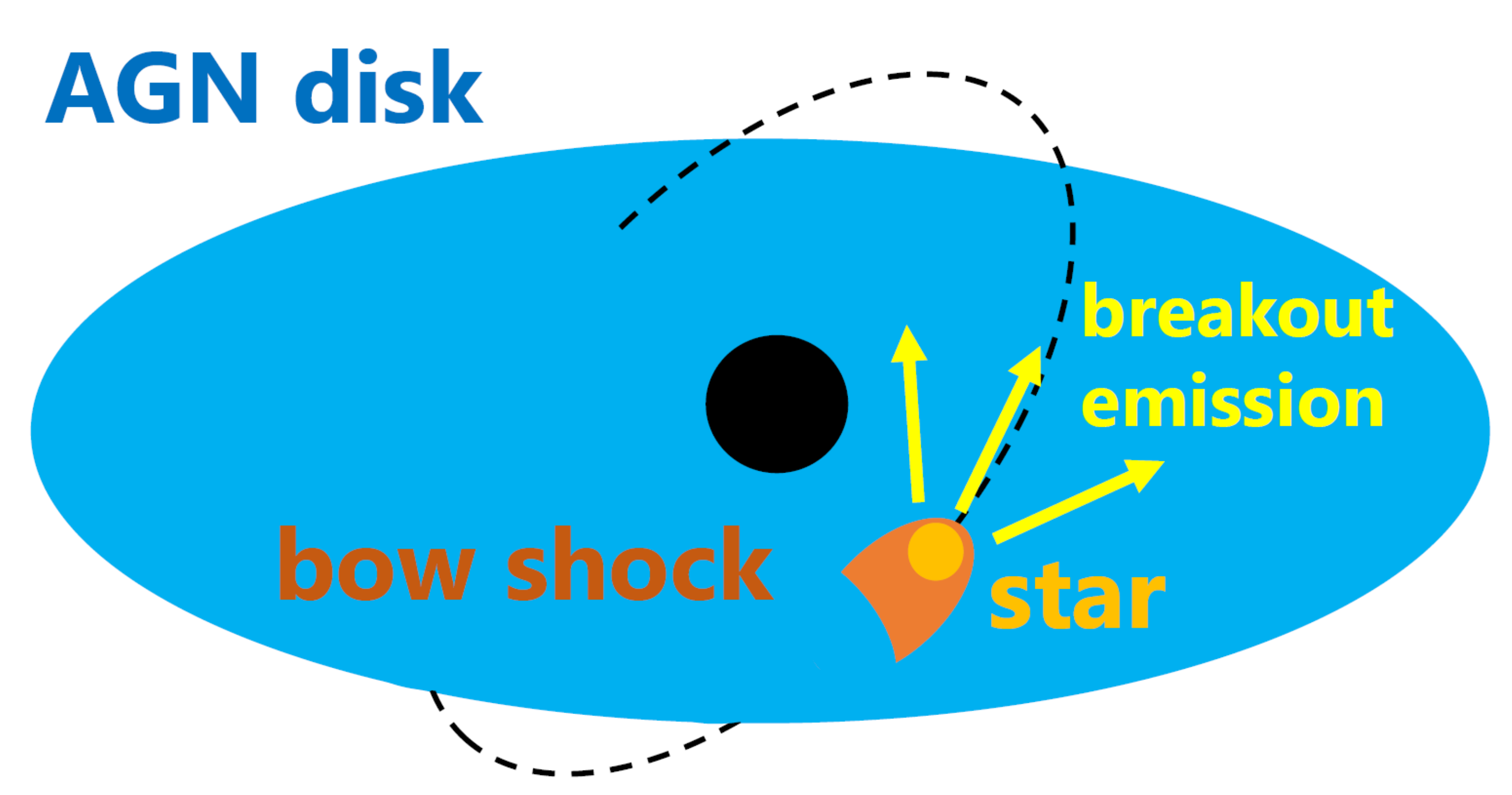}
\caption{
Schematic picture of the breakout emission from the collision between a star and an AGN disk.  
}
\label{fig:schematic_flare}
\end{center}
\end{figure}

\begin{table*}
\begin{center}
\caption{The properties of the observed QPE events. 
We refer to \citet{Miniutti2019}, 
\citet{Giustini2020}, 
\citet{Arcodia2021}, 
and \citet{Chakraborty2021} 
for the properties of GSN069, RXJ1301.9, eRO-QPE1 and eRO-QPE2, and XMMSL1, respectively. 
For the SMBH masses of eRO-QPE1 and eRO-QPE2, 
we refer to \citet{Chen2022}. 
}
\label{table:properties_events}
\begin{tabular}{c|c|c|c|c}
\hline 
The name & The SMBH mass$~[\Msun]$
& Period$~[\rm s]$
& Peak luminosity$~[\rm erg/s]$&Duration$~[\rm h]$\\
\hline\hline
GSN 069 & $4\times 10^5$ & $3.16\times 10^4$
&$5\times 10^{42}$&
0.5
\\\hline
RXJ1301.9 & $1.8\times 10^6$ & $1.65\times 10^4$
&$1\times 10^{42}$&$\sim 0.3$\\\hline
eRO-QPE1 & $9.1\times 10^5$ & $6.66\times 10^4$ 
&
$\sim 10^{43}$
&$7.6$\\\hline
eRO-QPE2 & $2.3\times 10^5$ & $8.6\times 10^3$
&$1\times 10^{42}$&$0.45$\\\hline
XMMSL1 & $8.5\times 10^4$ & $9.0\times 10^3$
&$3\times 10^{41}$&$\sim 0.3$\\\hline
\end{tabular}
\end{center}
\end{table*}

\section{Method}

In this section, 
we describe procedures to 
evaluate properties of the shocks and emission associated with the collisions between a star and an AGN disk. 
Readers not interested in the model details may skip directly to the next section, describing our results and the comparison to observed QPE properties.

\subsection{Model}

\label{sec:model}

\begin{figure}
\begin{center}
\includegraphics[width=85mm]{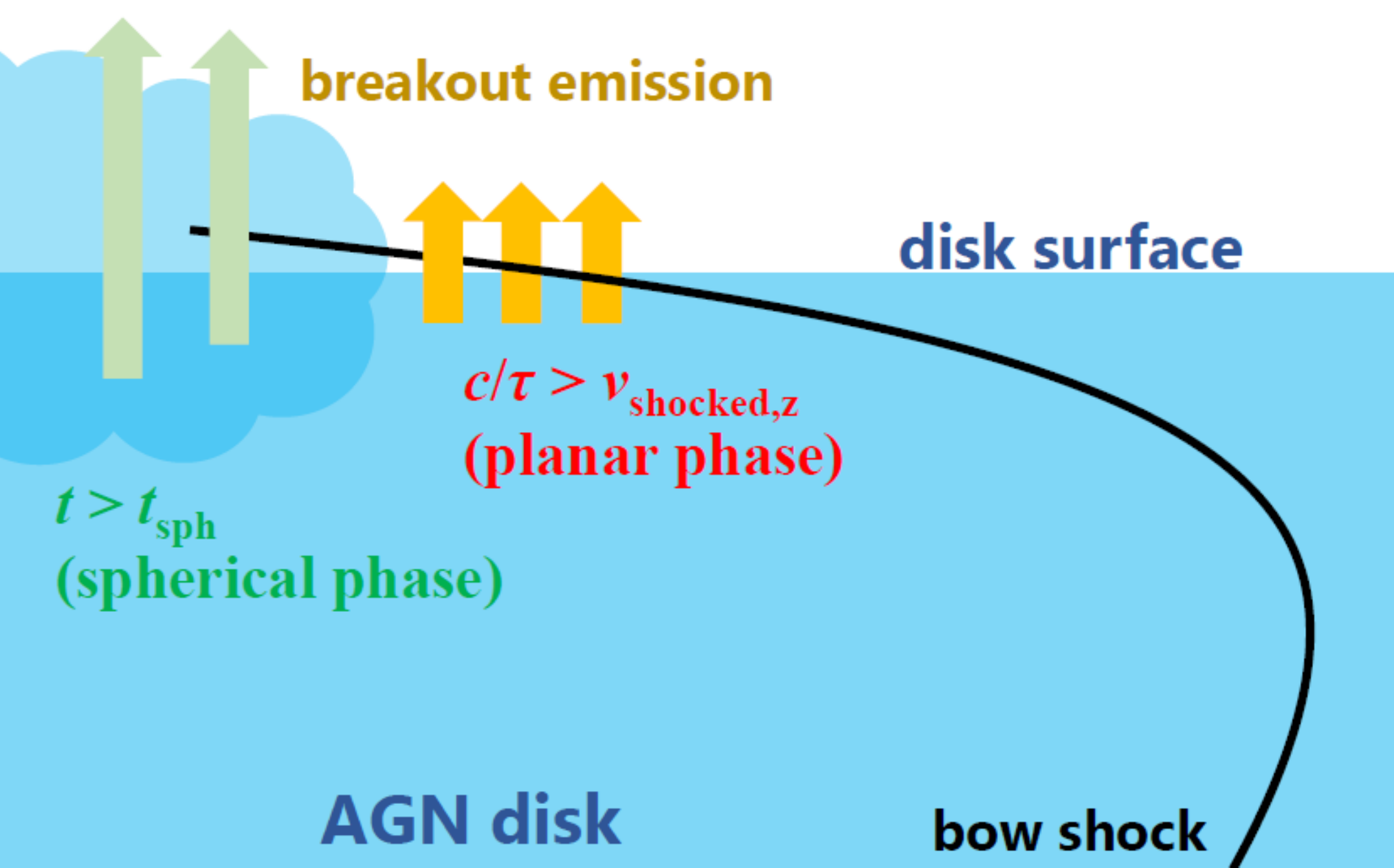}
\caption{
Schematic picture representing the breakout emission from the bow shock in the "planar" and "spherical" phases, corresponding to before and after the hot
shocked gas, emerging from the AGN disk surface, doubles its radius.
}
\label{fig:schematic_emission}
\end{center}
\end{figure}

First, we give a brief overview the emission model. 
When a star moves at a supersonic speed within the AGN disk, a bow shock forms along a conical region around and behind the star,
if the ram pressure in the star's envelope or the stellar wind is higher than the pressure of the disk (Fig.~\ref{fig:schematic_flare}).
We assume that the shape of this shock surface is given by \citet{Wilkin1996}. 
The conical region contains hot post-shock gas, moving along with the star.
If the relative velocity of the star to the AGN disk gas is high enough, as expected for stars orbiting around the SMBH with 
the short orbital time corresponding to 
the recurrence periods of the QPE events (Table~\ref{table:properties_events}), 
the thermal energy of the shocked gas is dominated by radiation. 
At any given time, 
at the intersection of this cone with the surface of the AGN disk, thermal photons from the hot gas break out once the diffusion speed of photons becomes faster than the shock velocity (Fig.~\ref{fig:schematic_emission}). 
To calculate the properties of the emission, we simply sum the emission from each patch along this intersection. 
Since the duration of the breakout emission from each patch is short, 
a low-inclination orbit for the star is found to be required 
to increase the total area from which the breakout emission is released at any given time.

In order to compute the emerging luminosity, we need to describe the motion of the star in the AGN disk, and the geometry of the conical bow-shock as it intersects the AGN disk surface. We assume that the breakout occurs at the AGN density of $\rho_{\rm AGN}$ and the height from the AGN disk mid-plane of $z=H_{\rm AGN}$. 
We erect Cartesian coordinates $(x,y,z)$, 
with the origin at the position of the star 
in the midplane of the AGN disk, the $z$-axis set to the angular momentum  direction of the AGN disk, and the $x$-axis chosen so that the star moves  in the $x-z$ plane. 
We refer to the components of vectors in these 
coordinates by the subscripts $x$, $y$, or $z$. 
The vertical gas density profile around $z=H_{\rm AGN}$ is assumed to be described by $\rho=\rho_{\rm AGN} (d/d_0)^n$, where $d$ is the distance to the edge of the AGN disk, $d_0$ is $d$ at which photons begin to break out the AGN disk, 
and we assume $n=6$ in our fiducial model, referring to the density profile of radiation pressure dominated disks \citep{Grishin2021}.

\begin{figure}
\begin{center}
\includegraphics[width=85mm]{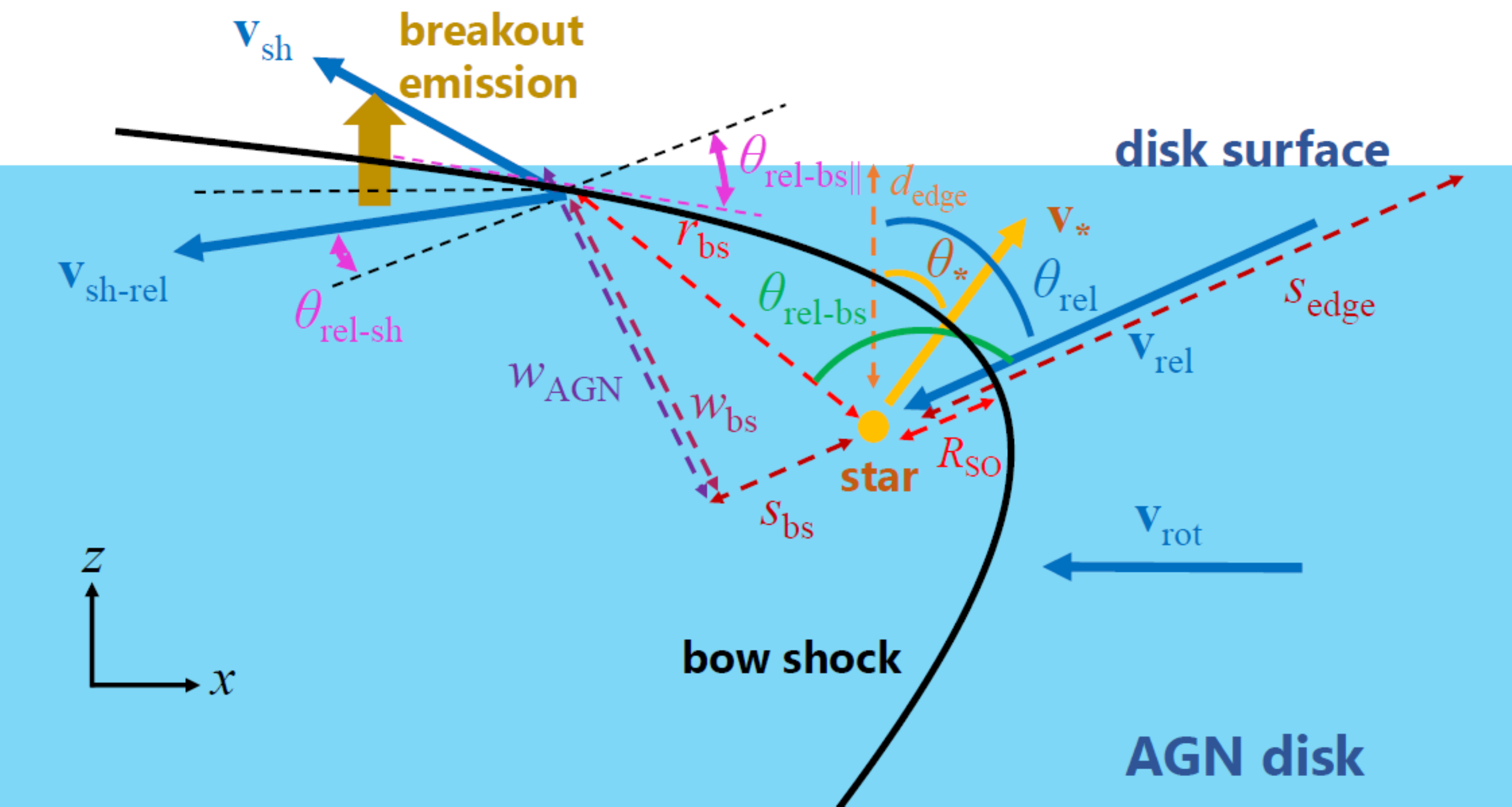}
\caption{
The geometry of the bow shock and the various distances, velocities and angles as defined in our coordinate systems. The star is moving through the AGN disk in the $y=0$ plane. The quantities shown in the figure are defined in the rest frame of the star, except for the velocities ${\bf v}_*$ and ${\bf v}_{\rm sh}$ which are depicted in the rest frame of the SMBH. 
}
\label{fig:schematic_configuration}
\end{center}
\end{figure}

For simplicity, we assume that 
the star is moving across the AGN disk along a straight trajectory. 
The stellar velocity is given as ${\bf v}_* =v_*[{\rm sin}(\theta_*), 0, {\rm cos}(\theta_*)]$, and the stellar position is ${\bf x}_* =(x_*,y_*,z_*)=v_*t[{\rm sin}(\theta_*), 0, {\rm cos}(\theta_*)]$, where 
$\theta_*$ is the zenith angle between the orbital direction of the star and the angular momentum direction of the AGN disk, 
$t$ is time, and $t=0$ is set to the time at which the star is in the disk midplane ($z_*=0$).  The speed of the star is
$v_*=f_e v_{\rm Kep}$, where
$v_{\rm Kep}=(GM_{\rm SMBH}/r)^{1/2}$ is the Keplerian velocity at the distance $r$ from the SMBH and
$f_e$ is a factor ranging from $1-e$ to $1+e$ between the apocentre and the pericentre of the orbit, 
and $e$ is the orbital eccentricity of the star. 
We set the AGN gas velocity to ${\bf v}_{\rm rot} = v_{\rm Kep}(-1,0,0)$, 
and then, the relative velocity of local gas with respect to the stellar motion is given by 
${\bf v}_{\rm rel}={\bf v}_{\rm rot}-{\bf v}_{\rm *}$, 
and the zenith angle of the direction of ${\bf v}_{\rm rel}$ with respect to the angular momentum direction of the AGN disk is given by $\theta_{\rm rel}={\rm arctan}(v_{{\rm rel},x}/v_{{\rm rel},z})$. See Fig.~\ref{fig:schematic_configuration} for illustrating these quantities.

\begin{figure}
\begin{center}
\includegraphics[width=65mm]{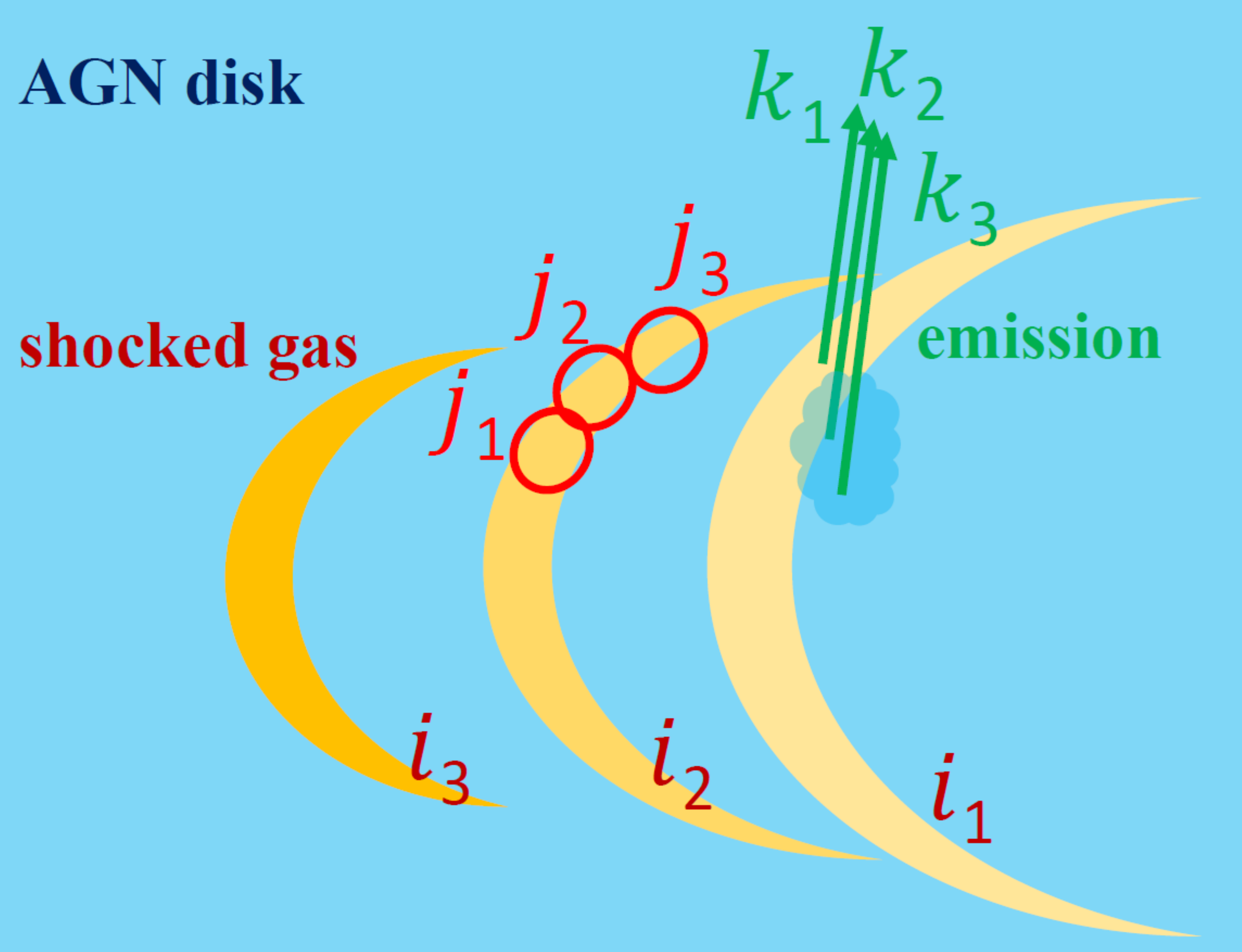}
\caption{
Schematic diagram representing the construction of the light-curve from the system. At any given instant (labeled by $i$), the intersection of the bow-shock with the disk surface defines a 1D curve (yellow), trailing the moving star.  Each patch along this break-out curve (labeled by $j$) begins to produce a light-curve.   At any given time (labeled by $k$) we observe this system, we sum the contribution from each patch along the break-out curve at each previous position of the star. 
}
\label{fig:schematic_ijk}
\end{center}
\end{figure}

Next, we consider the geometry of the bow shocks. 
Referring to previous studies 
\citep{Wilkin1996,Mohamed2012}, 
we assume that the asymptotic shape of a bow shock around the star, with the stand-off radius for the head-on stream $R_{\rm SO}$ in the stellar rest frame, is given by 
\begin{eqnarray}
\label{eq:r_sh}
r_{\rm bs}/R_{\rm SO}=(3-3\theta_{\rm rel-bs}/{\rm tan}\theta_{\rm rel-bs})^{1/2}/{\rm sin}(\theta_{\rm rel-bs}),
\end{eqnarray}
where $r_{\rm bs}$ is the bow-shock distance from the star, and $\theta_{\rm rel-bs}$ is the angle between ${\bf v}_{\rm rel}$ and the direction of the bow shock with respect to the star. For a less massive star, the ram pressure of cold gas is equated to the pressure of the envelope of the star, and hence, the distance to the shock in the head-on direction ($R_{\rm SO}$) is roughly given by the radius of the star. This can be different for a massive star, for which the ram pressure of the stellar wind can be equated to that of the unshocked gas at a much larger radius. 
Note that the shapes of the bow shocks formed by a solid surface assumed in our model and those produced in the presence of winds considered in \citealt{Wilkin1996} are almost identical, as shown by numerical simulations in \citet{Yalinewich2016}. 

In the rest of this section, we describe variables in the rest frame of the star unless stated otherwise. 
The projected distance to the bow shock at $r_{\rm bs}$ (and $\theta_{\rm rel-bs}$) from the star along
the trajectory of the cold gas motion 
is $s_{\rm bs}=r_{\rm bs}{\rm cos}(\pi-\theta_{\rm rel-bs})$, 
and the half-shock width 
is $w_{\rm bs}=r_{\rm bs}{\rm sin}(\pi-\theta_{\rm rel-bs})$. 
At each position of the star, 
the bow shock breaks out of the AGN surface 
in the direction 
in which the width of the bow shock $w_{\rm bs}$ is equal to 
the distance from the trajectory of unshocked gas 
across the star to 
the position at which breakout occurs 
$w_{\rm AGN}=(s_{\rm bs}+s_{\rm edge})/[{\rm tan}(\theta_{\rm rel}){\rm cos}\phi_{y}]$, 
where 
$s_{\rm edge}=d_{\rm edge}/{\rm cos}(\theta_{\rm rel})$ is the distance of the star to the surface of the AGN disk in the direction of the unshocked gas motion, 
$d_{\rm edge}=H_{\rm AGN}-z_*$ 
is the distance from the star to the height at which breakout occurs, 
$\phi_y\equiv {\rm arctan}(y_{\rm bs}/w_{\rm bs})$, 
and $y_{\rm bs}$ is the $y$-coordinate of the bow shock.

The velocities of the unshocked gas parallel and perpendicular to the bow shock surface 
are, respectively, 
$v_{\rm c,||}=v_{\rm rel} {\rm cos} (\theta_{\rm rel-bs||})$ and 
$v_{\rm c,\perp}=v_{\rm rel} {\rm sin} (\theta_{\rm rel-bs||})$, 
where $\theta_{\rm rel-bs||}={\rm arctan}({\rm d}w_{\rm bs}/{\rm d}s_{\rm bs})$ is 
the angle between the bow shock surface at $r_{\rm bs}$ (and $\theta_{\rm rel-bs}$) and the unshocked gas motion. 
Through the bow shock, 
the velocities of the shocked gas parallel and perpendicular to the shock surface 
are, respectively, related to the velocity of the cold (unshocked) gas as 
$v_{\rm sh,||}=v_{\rm c,||}$ and 
$v_{\rm sh,\perp}=v_{\rm c,\perp}(\gamma-1/\gamma+1)$, 
and the sound speed of the shocked gas is given as 
$c_{\rm s}=\sqrt{2\gamma/(\gamma-1)} v_{\rm sh,\perp}$ 
assuming strong shocks. Here $\gamma$ is the adiabatic index, which we set to $\gamma=4/3$ as gas is dominated by radiation pressure. 
The angle between the shocked and unshocked gas motions 
is 
\begin{eqnarray}
\label{eq:th_sh_cold}
\theta_{\rm rel-sh}={\rm atan}(v_{\rm sh,\perp}/v_{\rm sh,||}).
\end{eqnarray}
The three-dimensional shocked gas velocity in the stellar rest frame (${\bf v}_{\rm sh-rel}$) can be derived by considering the fact that the shocked gas motion is directed to the $-x'''$-direction, where ${\bf x}'''$ is rotated from ${\bf x}''$ around the $y''$-axis by $\theta_{\rm rel-sh}$, ${\bf x}''$ is rotated from ${\bf x}'$ around the $x'$-axis by $-\phi_y$, and ${\bf x}'$ is rotated from ${\bf x}$ around the $y$-axis by $\theta_{\rm rel}-\pi/2$. 
Since the velocities above are estimated in the rest frame of the star, 
the velocity of the shocked gas in the rest frame of the central SMBH is given by 
\begin{eqnarray}
\label{eq:v_sh}
{\bf v}_{\rm sh}={\bf v}_{\rm sh-rel}+{\bf v}_*. 
\end{eqnarray}

To numerically evaluate the properties of the shocks and the emission, both of which depend on the stellar position, the 
shape and orientation of the bow shock,
and the time from the breakout, 
we discretize the position, the 
patches along the bow shock,
and the time from the breakout by $i_{\rm max}^{\rm th}$, $j_{\rm max}^{\rm th}$, and $k_{\rm max}^{\rm th}$ cells, respectively, with 
$i_{\rm max}=j_{\rm max}=k_{\rm max}=500$ 
(Fig.~\ref{fig:schematic_ijk}). 
First, we discretize the stellar position as 
${\bf x}_*={\bf x}_i=v_*t_i[{\rm sin}(\theta_*),0,{\rm cos}(\theta_*)]$, 
where 
$t_i=i\Delta t$ is the discretized time, 
$i$ is a positive integer, 
$\Delta t =t_{\rm fin}/i_{\rm max}$ is the timestep, 
and we set the final time $t_{\rm fin}$ to be $t_{\rm max}=10H_{\rm AGN}/v_{*,z}$ to ensure that the star's trajectory is followed until it goes well outside the AGN disk.

At each position of the star, the surface of the bow shock 
is 
described 
by Eq.~\eqref{eq:r_sh}, and thus, the 
regions of the shocks breaking out of the AGN disk can be calculated. 
These regions are defined by the intersection of the two-dimensional parabolic bow shock surface with the flat AGN disk surface -- at a given time $i$, they correspond to a one-dimensional curve. 
For the star at the $i$th position, 
points along this break-out curve can be enumerated 
using the angle between 
the direction from the star toward the break-out point and the star's velocity vector, 
$\theta_{{\rm rel-bs,BO},i,j}=\theta_{{\rm rel-bs,BO},i,y0}+(\pi-\theta_{{\rm rel-bs,BO},i,y0})(j/j_{\rm max})$, 
where 
$\theta_{{\rm rel-bs,BO},i,y0}$ is $\theta_{{\rm rel-bs}}$ for the 
break-out point in the 
$y_{\rm bs}=0$ 
plane (see the green angle in Fig.~\ref{fig:schematic_configuration}) 
and $j$ is a positive integer. 
Note that the angle is smallest in the $y_{\rm bs}=0$ plane and approaches $\pi$ at infinite distance from the star. 

Below we designate the physical quantities of the shocks at $\theta_{{\rm rel-bs,BO},i,j}$ 
with the subscripts $i,j$. 
We calculate the surface area of the $j^{\rm th}$ shocked region when the star is at ${\bf x}_*={\bf x}_i$ as
$S_{i,j}=\Delta x \Delta y_{i,j}$
where 
$\Delta x =v_{{\rm rel},x}\Delta t$ is the length that the breakout region moves in the $x$-direction within $\Delta t$ due to the motion of the star, 
$\Delta y_{i,j}$ and $l_{i,j}$ are the $y$-direction length and the length of the shock breaking out the AGN disk between the $j^{\rm th}$--$(j+1)^{\rm th}$ cell at $t=t_i$. 
$\Delta y_{i,j}=y_{i,j+1}-y_{i,j}$ is derived from the relation 
\begin{eqnarray}
\label{eq:y_ij}
-s_{\rm sh}{\rm cos}\theta_{\rm rel}+(w_{\rm sh}^2-y_{\rm sh}^2)^{1/2}{\rm cos}\theta_{\rm rel}+z_*=H_{\rm AGN}. 
\end{eqnarray} 
We only consider the emission from shocks for 
$\theta_{\rm rel-bs}>0$.
This is because the breakout emission for $\theta_{\rm rel-bs}<0$ (i.e. when the "bottom half" of the bow shock in Fig.~\ref{fig:schematic_configuration}, below the star, is breaking out) is likely covered by gas already shocked and expanding. 
Additionally, we assume that shocks emerge only when $c_{s,i,j}>c_{s, \rm AGN}$ since otherwise strong shocks are not expected. Here $c_{s,\rm AGN}$ is the sound speed in the unshocked AGN disk gas, and we assume $c_{s,\rm AGN}=H_{\rm AGN}v_{\rm Kep}/r$.

After breakout of each patch of the bow shock, 
we calculate the luminosity and temperature evolution 
following \citet{Nakar2010}. 
The emission from the area $S_{i,j}$ is 
\begin{eqnarray}
\label{eq:li_t}
&L_{i,j}(t-t_i)
\sim
S_{i,j} 
v_{{\rm c,\perp},i,j}^2\rho_{\rm AGN} 
v_{{\rm sh},z,i,j}\times \nonumber\\
&\left\{
\begin{array}{l}
0~~~~\mathrm{for}~~t-t_i\lesssim 0 ,\\
1~~~~\mathrm{for}~~0\lesssim t-t_i\lesssim t_{{\rm BO},i,j} ,\\
(t-t_i/t_{{\rm BO},i,j})^{-4/3}
~~~~\mathrm{for}~~t_{{\rm BO}i,j}\lesssim t-t_i\lesssim t_{{\rm sph},i,j} ,\\
(t_{{\rm sph},i,j}/t_{{\rm BO}i,j})^{-4/3}
(t-t_i/t_{{\rm sph},i,j})^{-\frac{2.28n-2}{3(1.19n+1)}}
~~~\mathrm{otherwise} ,
\end{array}
\right.
\end{eqnarray}
where $t_{{\rm BO},i,j}$ is the breakout timescale  on which the diffusion speed of photons becomes faster than the shock velocity, 
and $t_{{\rm sph},i,j}$ is the timescale at which the breakout region begins to expand in spherical directions 
for the shock at the $(i,j)^{\rm th}$ cell. 
In Eq.~\eqref{eq:li_t}, 
the first, second, and third row corresponds to the luminosity in the planar phase, and the fourth row corresponds to that in the spherical phase.
Here "planar" and "spherical" refer to phases before and after the
shocked gas doubles its radius through adiabatic expansion caused by the enhancement of its thermal energy, respectively. 
The evolution of the emission properties are significantly different during these two phases, 
mainly because the radius of the shocked envelope remains roughly constant during the planar phase, while it increases significantly during the spherically expanding phase.

The timescales can be calculated as $t_{{\rm BO},i,j}=c/(\kappa \rho_{\rm AGN} v_{{\rm sh},z,i,j}^2)$
and 
$t_{{\rm sph},i,j}=H_{\rm AGN}/ v_{{\rm sh}, z, i,j}$,
where $\kappa$ is the opacity of the AGN gas, and we set $\kappa=0.4~{\rm cm^{2}~g^{-1}}$ for electron scattering 
 (appropriate in the inner, ionised regions of AGN discs; see \S~\ref{sec:caveats} for other opacity regimes).  
We calculated the total luminosity at time $t_k$ as 
\begin{align}
\label{eq:ltot_t}
L_{\rm tot}(t_k)
=2\Sigma_i^{i_{\rm max}}  \Sigma_j^{j_{\rm max}} 
L_{i,j}(t_k), 
\end{align}
where $t_k=k\Delta t$ is the discretized time, 
$k$ is a positive integer, 
and the factor of 2 in Eq.~\eqref{eq:ltot_t} accounts for the assumption that the shock is symmetric with respect to the $y=0$ plane \footnote{In this assumption, we ignore the shear motion of the AGN gas for simplicity. }. 
During the transition between the two phases in Eq.~\eqref{eq:li_t}, 
we averaged the luminosities by weighting the duration spent in both phases between $t_k$--$t_{k+1}$.

The temperature of the breakout emission is also prescribed following \citet{Nakar2010}. 
Since the temperature evolution from $t_{{\rm sph},i,j}$ to the time at which the colour shell (whose emission characterizes the observed radiation temperature) reaches thermal equilibrium ($t_1$) is not explicitly given, 
we interpolate this range by a power law \footnote{This prescription negligibly affects the result of this paper, as the temperature evolution in this range is very quick ($\sim \propto t^{-5}$).}. 
After gas in the color shell becomes thermally coupled with radiation ($t_{2}$), we adopt the temperature dependence on the time given in the second line of Eq.~(17) in \citet{Nakar2010}. 

To compute the typical temperature of emission for the time $t=t_k$, we weigh the temperature at $i$, $j$, and $k$ ($T_{i,j,k}$) by the luminosity for $i$, $j$, and $k$ 
as 
\begin{align}
\label{eq:tem_typ}
T_{\rm typ}(t_k)
=2\Sigma_i^{i_{\rm max}}  \Sigma_j^{j_{\rm max}} 
L_{i,j}(t_k)T_{i,j,k}/
L_{\rm tot}(t_k).
\end{align}
To compute the X-ray luminosity, we assume black-body spectra with the temperature $T_{\rm typ}$, and the frequency of the X-ray ranges $0.4$--$3~{\rm keV}$, considering observations by XMM-Newton. Note that this prescription overestimates the X-ray luminosity when $k_{\rm B}T_{\rm typ}$ is much higher than that of the X-ray band and the radiation is out of the thermal equilibrium 
\footnote{In this case, the radiation is characterized by a Wien spectrum, $F_{\nu}\propto \nu^3 {\rm exp}(h\nu/k_{\rm B}T)$, where $h$ is the Planck constant and $\nu$ is the photon frequency.
}, 
but it is a good approximation in the fiducial model provided in the next section.

\begin{table}
\begin{center}
\caption{
Fiducial values of our model parameters. 
}
\label{table:parameter_fiducial}
\begin{tabular}{p{4.5cm}|c}
\hline 
Parameter & Fiducial value \\
\hline\hline
Stand-off radius for the bow shock& 
$R_{\rm SO}=3\times 10^{11}\,{\rm cm}$
\\\hline
Zenith angle between the orbital angular momentum direction of the star and the AGN disk & $\theta_{*}=80^\circ$\\\hline
AGN density& $\rho_{\rm AGN}=3\times 10^{-10}~{\rm g~cm^{-3}}$\\\hline
Scale height of the AGN disk& $H_{\rm AGN}=10^{12}~{\rm cm}$\\\hline
Orbital period of the star & $P_{\rm orb}=20\,{\rm hr}$\\\hline
SMBH mass & 
$M_{\rm SMBH}=10^6\,{\Msun}$
\\\hline
Orbital eccentricity of the star& $e=0$\\\hline
Power-law slope for the vertical AGN gas density profile& $n=6$\\\hline
\end{tabular}
\end{center}
\end{table}

\subsection{Initial condition}

\label{sec:initial_condition}

The various parameters of our model are summarised in 
Table~\ref{table:parameter_fiducial}. 
We assume that flares occur twice per orbit around the SMBH 
as the star collides with the AGN disk 
($2P_{\rm QPE}=P_{\rm orb}$, where $P_{\rm orb}$ is the orbital period of the star). 
In this case, the semi-major axis of the stellar orbit is estimated as 
$a=(P_{\rm QPE}/\pi)^{2/3}(GM_{\rm SMBH})^{1/3}$, where $P_{\rm QPE}$ is the period of QPEs. 
We adopt $e=0$ in the fiducial model as the orbital periods of QPEs are observed to be very regular (but see \citealt{Miniutti2022} and discussions below), 
and $M_{\rm SMBH}=10^6~\Msun$ 
and $P_{\rm QPE}=10~{\rm hr}$ reflecting the typical properties of QPEs (Table~\ref{table:properties_events}). 
We set the stand-off radius for the bow shock to 
$R_{\rm SO}=3\times 10^{11}~{\rm cm}$ 
assuming a star with a mass of $m_* \sim 10~\Msun$ \citep{Torres10}. 
We set the scale height of the AGN disk to $H_{\rm AGN}=10^{12}~{\rm cm}$, corresponding to the aspect ratio of $H_{\rm AGN}/a\sim 0.04 (P_{\rm QPE}/10~{\rm hr})^{-2/3}(M_{\rm SMBH}/10^6~\Msun)^{-1/3}$, 
and the AGN density to $\rho_{\rm AGN}=3\times 10^{-10}~{\rm g~cm^{-3}}$. 
The corresponding bolometric luminosity of the AGN disk is
\begin{eqnarray}
L_{\rm AGN}&\sim& 3\times 10^{42}~{\rm erg/s}
(H_{\rm AGN}/10^{12}~{\rm cm})^3 
(P_{\rm QPE}/10~{\rm hr})^{-1}
\nonumber\\
&&\times(\rho_{\rm AGN}/3\times 10^{-10}~{\rm g~cm^{-3}})
(\alpha/0.1)
(\eta/0.1), 
\end{eqnarray}
which is roughly consistent with the luminosity of the quiescent phases in QPEs 
assuming a standard disk model \citep{Lu2022,Linial2023}, 
and accretion onto the SMBH at $\sim 0.1$ times the Eddington accretion rate, 
where $\alpha$ is the viscosity parameter, and $\eta$ is the radiative efficiency. 
In our fiducial model, we assume that the star's orbit is retrograde  with respect to the AGN disk gas, since otherwise the emission is found to be much dimmer compared to that observed in the QPEs. 
The zenith angle between the orbital angular momentum directions 
of the star and the AGN disk ($\theta_*$) is set to $80^\circ$ to roughly reproduce the duration and the luminosity of QPEs.

\begin{figure*}
\begin{center}
\includegraphics[width=145mm]{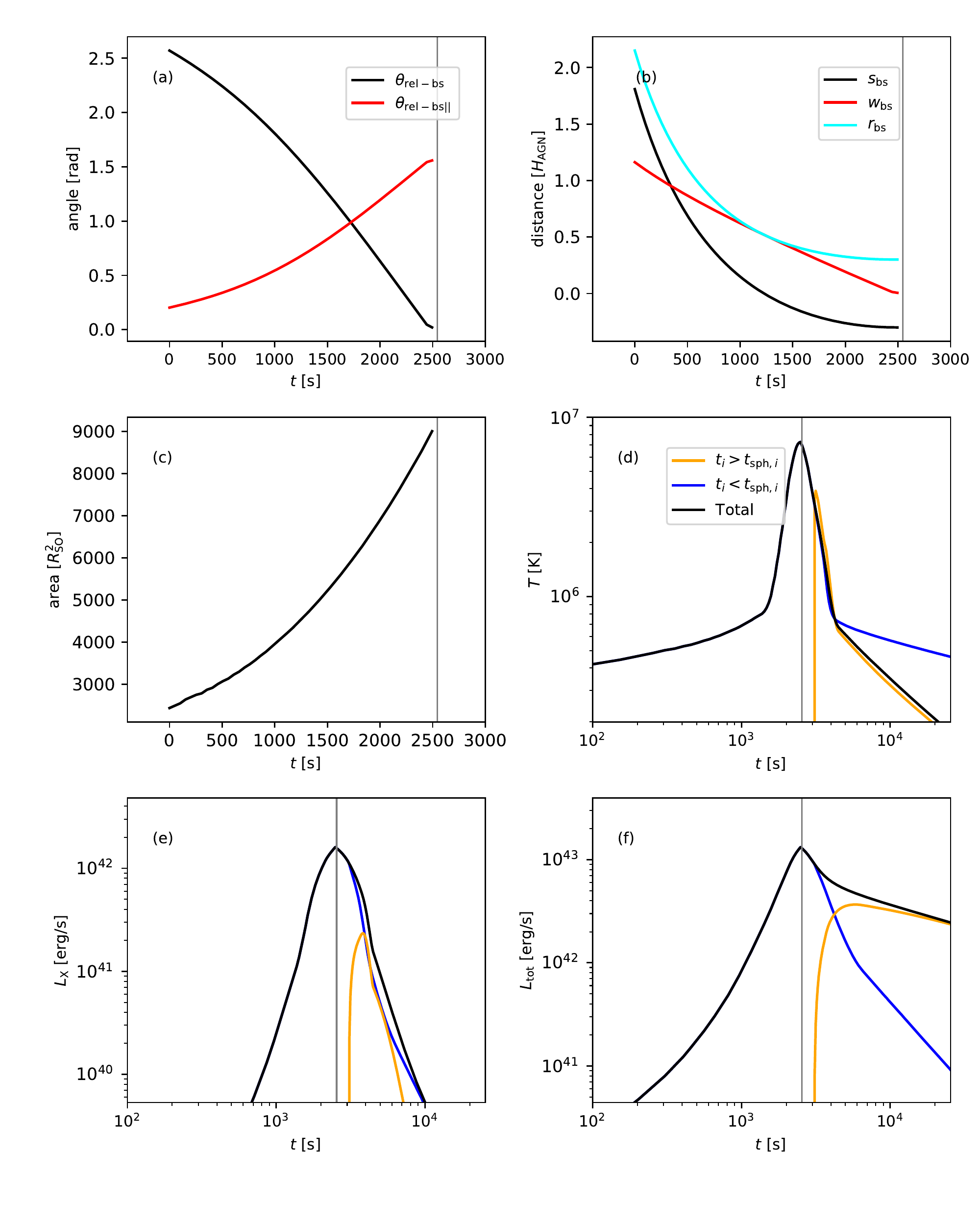}
\caption{
The evolution of the properties of the shock and the emission in the fiducial model. 
(a) The angle between the unshocked gas flow direction and the line connecting the point where the shock is breaking out of the disk and the star $\theta_{\rm rel-bs}$ (black) and the angle between the bow shock surface and the unshocked gas motion $\theta_{\rm rel-bs||}$ 
(red) 
on the plane along the stellar orbital motion and the angular momentum direction of the AGN ($y=0$) in the stellar rest frame.  
During the late stages, when the head of the bow-shock is exiting the AGN disk, the unshocked gas collides nearly perpendicularly with the surface of the bow shock, which strongly influences the properties (temperature and energy) of the shocked gas. 
(b) 
$w_{\rm bs}$ is the half-shock width 
(red), 
and $s_{\rm bs}$ (black) and $r_{\rm bs}$ 
(cyan) 
are, respectively, 
the projected distance along the trajectory of unshocked gas 
and the distance 
from the star to the bow shock breaking out of the AGN disk 
for $y=0$. 
These distances from the star to the shock breaking out of the disk all decrease with time. 
(c) The total area of the shocked gas 
emitting in the planar phase calculated as $\Sigma_j S_{i,j} \times t_{{\rm BO},i,j}/\Delta t$, in the units of the square of the stand-off radius ($R_{\rm SO}^2$), assumed to be equal to the stellar radius.
The total area emitting breakout emission is found to be much larger than the cross-sectional area of the star. 
(d) The luminosity-weighted temperature (colours are the same as in panel~e). 
(e) The X-ray luminosity between $0.4~{\rm keV}\leq h\nu \leq 3~{\rm KeV}$ (solid black). 
At any given time, the total luminosity is contributed by a sum of different patches of the bow-shock; some of these patches are in the planar phase, before they broke out of the AGN disk and started expanding ($t_i<t_{{\rm sph},i}$, blue), and some are afterwards, in a quasi-spherical post-breakout stage ($t_{{\rm sph},i}<t_i$, orange). 
(f) The bolometric luminosity (colours are the same as in panel~e). 
The vertical gray line marks the time at which the star exits the AGN disk surface. 
}
\label{fig:properties}
\end{center}
\end{figure*}

\begin{figure}
\begin{center}
\includegraphics[width=85mm]{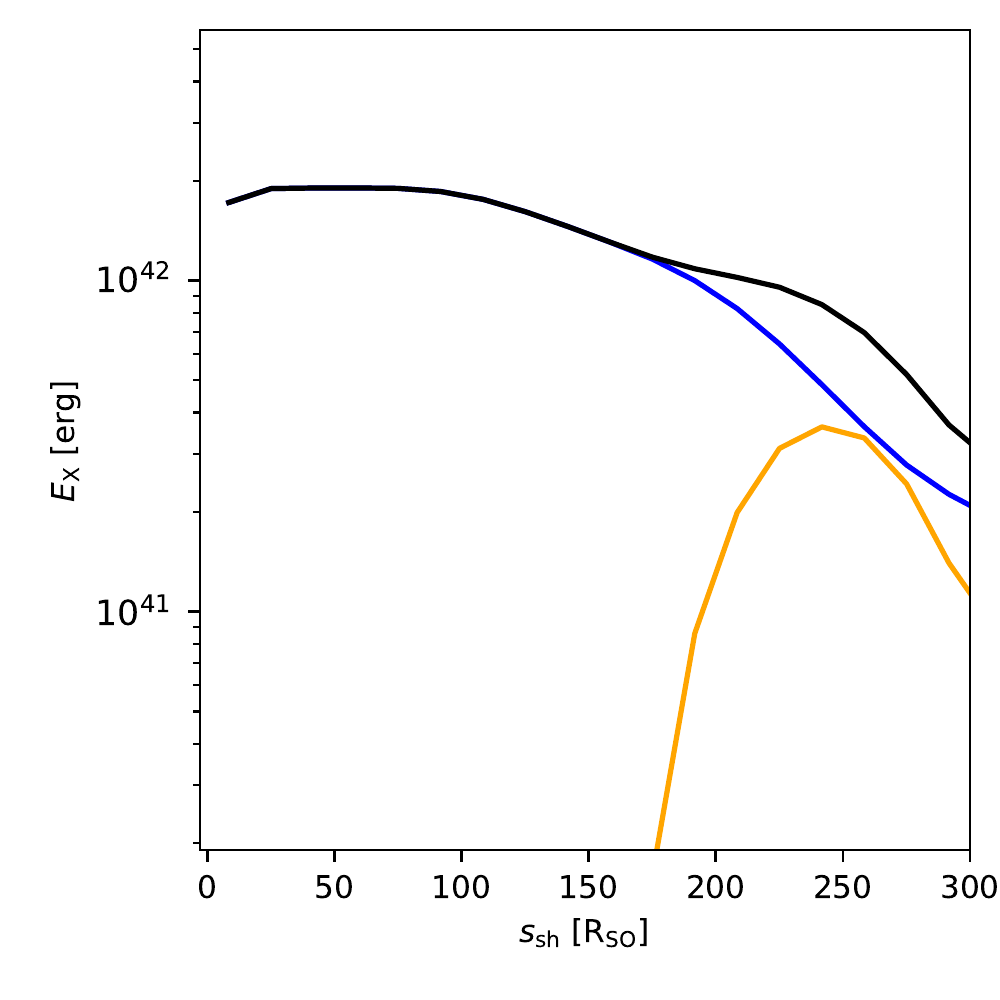}
\caption{
Contributions to the total X-ray luminosity integrated over the duration of the flare 
($E_{\rm x}(s_{{\rm bs},l})=\int L_{\rm x}(s_{{\rm bs},l})dt$, where $L_{\rm x}(s_{{\rm bs},l})$ is the X-ray luminosity from shocked regions in the $l$th cell $s_{\rm bs}=s_{{\rm bs},l}$, whose cell size is $1.8\times 10^9~{\rm cm}$) 
from 
shocked regions at different projected distances from the star, along the unshocked gas trajectory ($s_{\rm bs}$), shown in units of the stand-off radius ($R_{\rm SO}$). 
The black, blue, and orange lines are the total luminosity, the luminosity in the planar phase and that in the spherical phase, respectively. 
}
\label{fig:ex_s}
\end{center}
\end{figure}

\section{Results}

\subsection{Emission properties}

We first show the evolution of the bow shock and the corresponding emission. 
Fig.~\ref{fig:properties} 
represents the evolution of several properties. 
At $t=0$, the star is at the midplane of the AGN disk, 
and it passes out of the disk at 
$t=H_{\rm AGN}/v_*{\rm cos}(\theta_*)=2.6\times10^3~{\rm s}$. 
At the beginning of the calculation of $t=0$, 
the shocks break out from the AGN disk at the far side,
where the angle between the unshocked gas motion and the bow shock direction from the star is
$\theta_{\rm rel-bs}=2.6~{\rm rad}$ 
(black line in panel~a) and
the projected distance to the shock from the star along trajectory of the cold gas motion is
$s_{\rm bs}\sim 1.8~H_{\rm AGN}$ 
(black line in panel~b). 
The direction of unshocked gas flow 
is almost parallel to the shock surface with $\theta_{\rm rel-bs||}=0.20~{\rm rad}$ 
(red line in panel~a), 
and the sound velocity of the shocked gas breaking out of the AGN disk at $y=0$ 
is 
$c_s=0.01c$. 
The surface area releasing breakout emission in the planar phase 
is much larger than the cross-sectional area of the star (panel~c) due to the low inclination of the orbit. 

As the star approaches the surface of the AGN disk 
($r_{\rm bs}$ decreases, 
cyan line in panel~b), 
$\theta_{\rm rel-bs||}$ 
monotonically increases, and accordingly $c_s$ at $y=0$ increases to 
$c_s=0.06c$ 
since upstream gas flows more vertically onto the shock surface. 
The surface area of the regions breaking out of the AGN disk also increases (panel~c), since the condition for shocks ($c_{s,i,j}>c_{s,\rm AGN}$) is satisfied in large areas in the later phases. 
In $t> 2.6\times 10^3~{\rm s}$, 
$\theta_{\rm rel-bs}$ becomes negative, and hence, the emission from the shocks formed after this phase is not considered in this paper as stated in \S~\ref{sec:model}.

After the star passes through the AGN disk surface at 
$t\sim 2.6\times 10^3~{\rm s}$ 
(dashed gray lines in panels~d, e, f), 
the temperature of the shocked gas decreases with time (black line in panel~d). 
Also, in the planar phase, the radiation is out of thermal equilibrium as photons are deficient and the radiation temperature is higher than the black-body temperature due to the Comptonization of photons. On the other hand, at the beginning of the spherical phase, the radiation temperature rapidly decreases as the radiation reaches thermal equilibrium. 
Due to the temperature evolution, the X-ray luminosity also rapidly decreases (black line in panel~e). 
On the other hand, the bolometric luminosity roughly evolves following $\propto t^{-4/3}$ in the planar phase and $\propto t^{-0.5}$ in the spherical phase \citep{Nakar2010}. 
Note that most of the emission is contributed by 
patches along the bow shock at distances
far beyond
the stand-off radius 
($w_{\rm bs}>R_{\rm SO}$). 
In this model, the contribution from the emission in the spherical phase to the X-ray luminosity is minor compared to that in the planar phase (blue and orange lines in panel~e). 
Cases with more complicated light curves are shown as variations from our fiducial model in the next section.

In this fiducial model, 
the maximum luminosity in the X-ray band is 
$L_{\rm x,max}=1.6\times 10^{42}~{\rm erg/s}$, the duration for which the X-ray luminosity is above $0.1~L_{\rm x,max}$ is $3.0\times 10^3~{\rm s}$, the total radiated bolometric energy is $9.4\times 10^{46}~{\rm erg}$, and that in the X-ray band is $2.8\times 10^{45}~{\rm erg}$. 
Regions whose half-shock width at breakout is $w_{\rm bs}=$0--2, 2--3, and $3$--$5~R_{\rm SO}$ contribute 10, 82, and $8\%$ of 
the X-ray luminosity, respectively. 
Also, $59\%$ and $41\%$ 
of the X-ray luminosity is contributed by the shocks breaking out of the disk at $t<2500~{\rm s}$ and $t>2500~{\rm s}$, respectively. 
Since the duration for the breakout emission ($\sim 10^3~{\rm s}$ for the planar phase) is not short compared to the dynamical time of the stellar orbit in the AGN disk, 
the shocked gas keeps emitting until its distance from the star becomes much larger than the size of the stand-off radius (Fig.~\ref{fig:ex_s}).  
Hence, the X-ray luminosity is contributed mostly by early shocks formed in the regions with 
$w_{\rm bs}=2$--$3~R_{\rm SO}$, 
and this emission is released far from the star of $s_{\rm bs}\sim 100~R_{\rm SO}$. 
The values of the X-ray luminosity, 
the duration of the X-ray flare, and the temperature are consistent with those typically observed in QPEs (Table~\ref{table:properties_events}). 
Additionally, the rapid decrease in the temperature is consistent with the trend in the evolution of the hardness ratio found in \citet{Miniutti2022}.

\begin{figure*}
\begin{center}
\includegraphics[width=155mm]{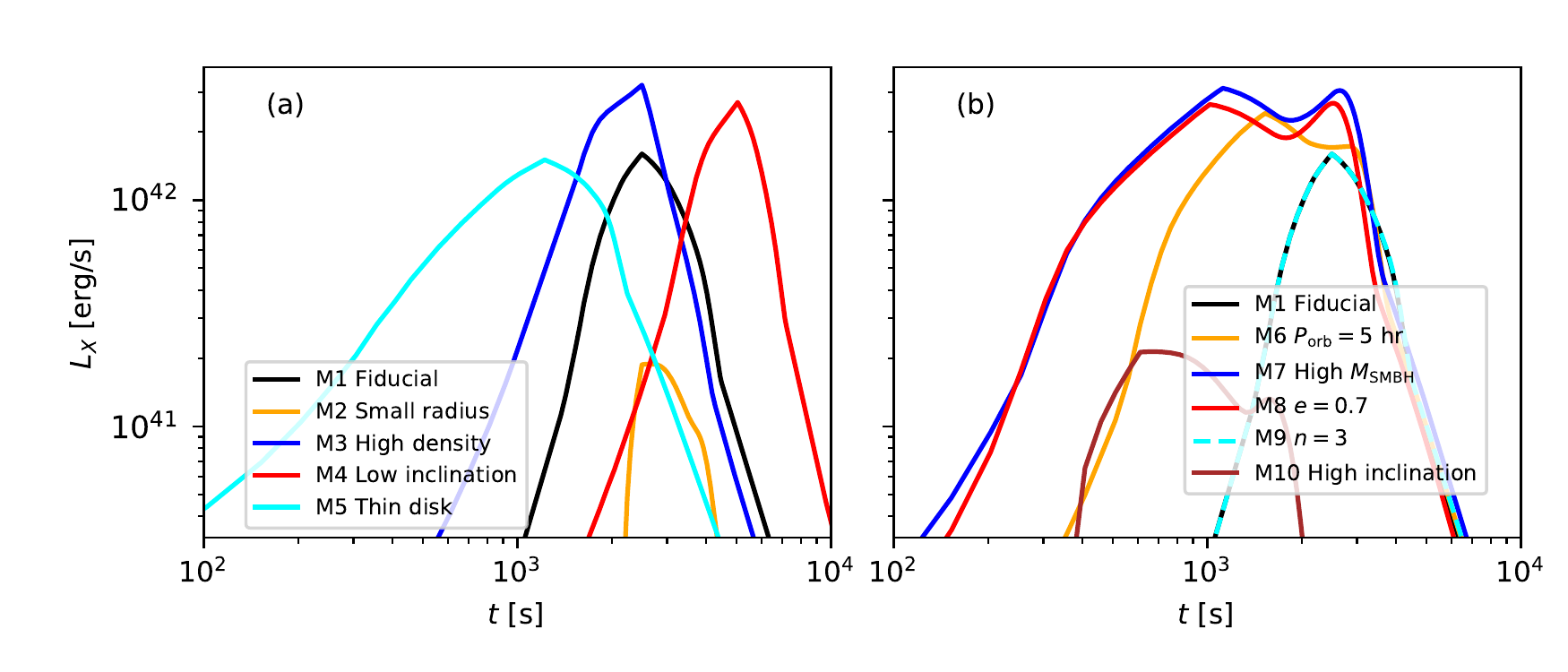}
\caption{
The parameter dependence of the X-ray light curves. 
(a) The results for the fiducial model (black), 
for 
$R_{\rm SO}=10^{11}~{\rm cm}$ (orange), 
$\rho_{\rm AGN}=10^{-9}~{\rm g~cm^{-3}}$ (blue), 
$\theta_*=85^\circ$ (red), 
and $H_{\rm AGN}=5\times 10^{11}~{\rm cm}$ (cyan). 
(b) The results for the same fiducial model (black), 
$P_{\rm orb}=5~{\rm hr}$ (orange), 
$M_{\rm SMBH}=10^7~\Msun$ (blue), 
$e=0.7$ (red), 
$n=3$ (cyan), 
and $\theta_*=50^\circ$ (brown). 
}
\label{fig:light_curves}
\end{center}
\end{figure*}

\begin{figure}
\begin{center}
\includegraphics[width=85mm]{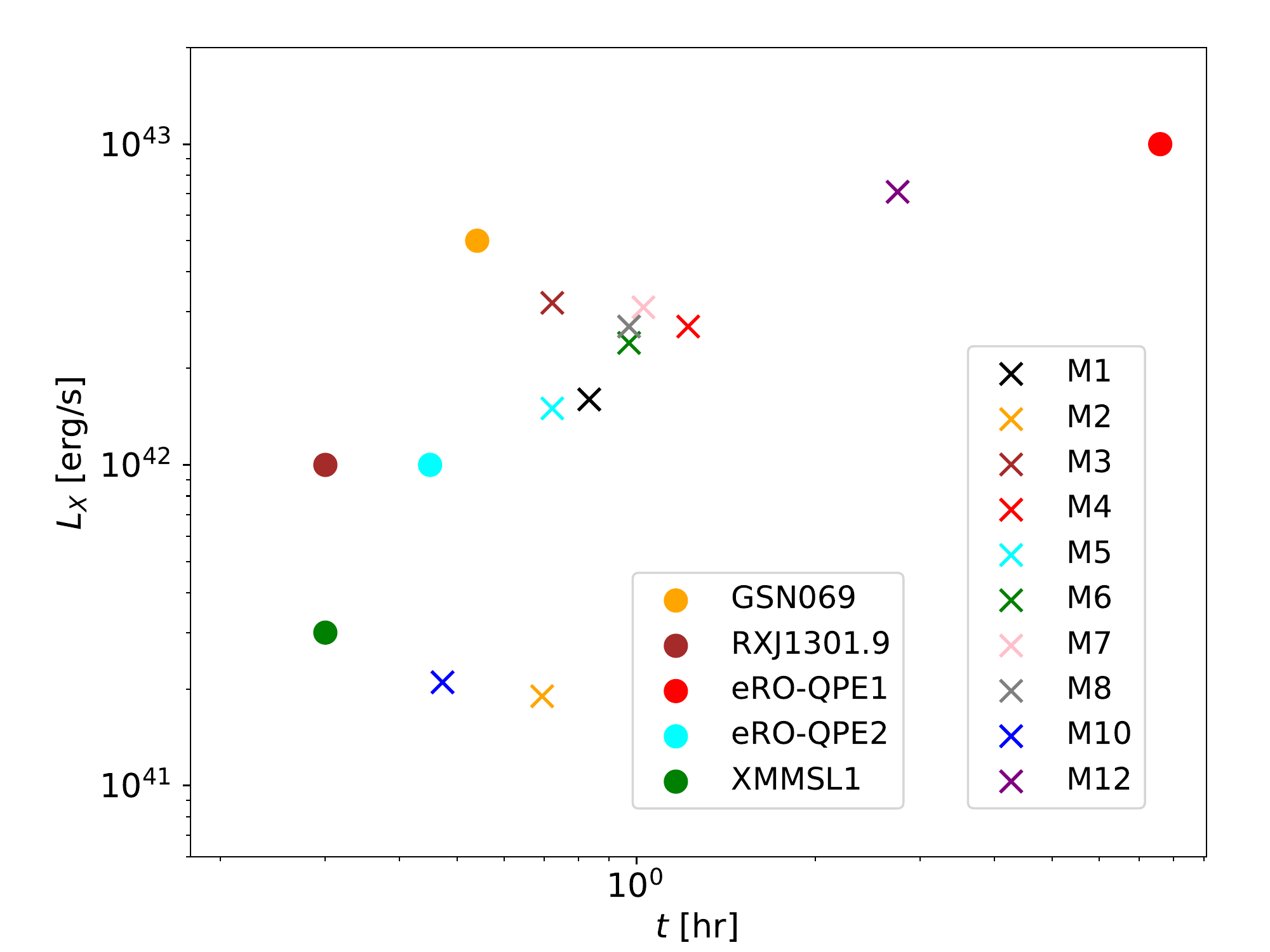}
\caption{
The distribution of the maximum value of $L_{\rm X}$ and the flare duration 
for the five observed events (filled circles) 
and 
models~M1--M8, M10, and M12 (crosses). 
}
\label{fig:l_t}
\end{center}
\end{figure}

\begin{table*}
\begin{center}
\caption{
Results in several different model-variants. 
The first column indicates the model number. 
In the “parameter”
column, we indicate parameters that deviate from the fiducial model. 
}
\label{table:telescope}
\hspace{-5mm}
\begin{tabular}{c|c|c|c|c|c}
\hline 
Model&Parameter
& Maximum $L_{\rm x}~[{\rm erg/s}]$& Duration~$[\rm s]$& X-ray energy $[\rm erg]$& bolometric energy $[\rm erg]$\\
\hline\hline
M1&Fiducial&$1.6\times 10^{42}$&$3.0\times 10^3$&
$3\times 10^{45}$&
$9\times 10^{46}$\\\hline

M2&
Smaller radius ($R_{\rm SO}=10^{11}$)
&
$1.9\times 10^{41}$&$2.5\times 10^3$&$3\times 10^{44}$&$1\times 10^{46}$\\\hline

M3&Higher gas density ($\rho_{\rm AGN}=10^{-9}~{\rm g~cm^{-3}}$)&$3.2\times 10^{42}$&$2.6\times 10^3$&
$5\times 10^{45}$&
$1\times 10^{47}$\\\hline

M4&Less inclined orbit ($\theta_*=85^\circ$)&
$2.7\times 10^{42}$&$4.4\times 10^3$&
$7\times 10^{45}$&
$2\times 10^{47}$\\\hline

M5&A thinner AGN disk ($H_{\rm AGN}=5\times 10^{11}~{\rm cm}$)&
$1.5\times 10^{42}$&$2.6\times 10^3$&
$2\times 10^{45}$&
$1\times 10^{47}$\\\hline

M6&
Shorter orbital period 
($P_{\rm orb}=5~{\rm hr}$)
&$2.4\times 10^{42}$&$3.5\times 10^3$&
$5\times 10^{45}$&
$2\times 10^{47}$\\\hline

M7&
Massive SMBH mass 
($M_{\rm SMBH}=10^7$)
&
$3.1\times 10^{42}$&$3.7\times 10^3$&
$8\times 10^{45}$&
$3\times 10^{47}$\\\hline

M8&
High eccentricity 
($e=0.7$)
&
$2.7\times 10^{42}$&$3.5\times 10^3$&
$6\times 10^{45}$&
$2\times 10^{47}$\\\hline

M9&Shallower vertical density gradient ($n=3$)&$1.6\times 10^{42}$&$3.1\times 10^3$&
$3\times 10^{45}$&
$1\times 10^{47}$\\\hline

M10
&Highly inclined orbit ($\theta_*=50^\circ$)&$2.1\times 10^{41}$&$1.7\times 10^3$&$2\times 10^{44}$&$1\times 10^{46}$\\\hline

M11
&Highly inclined orbit ($\theta_*=0^\circ$)&
$1.2\times 10^{40}$&$1.2\times 10^3$&
$9\times 10^{42}$&
$1\times 10^{44}$\\\hline

M12
&eRO-QPE1-like ($R_{\rm SO}=6\times 10^{11}, \theta_*=87^\circ$)&$7.1\times 10^{42}$&$9.9\times 10^3$&
$4\times 10^{46}$&
$2\times 10^{48}$\\\hline

\end{tabular}
\end{center}
\end{table*}

\subsection{Parameter dependence}

Next, we investigate the parameter dependence of the light curve in the X-ray band. 
We change the following parameters from the fiducial model ($\S$~\ref{sec:initial_condition}): 

\noindent -- 
in model~M2, 
a smaller bow shock size of  
$R_{\rm SO}=1\times 10^{11}~{\rm cm}$
to investigate the dependence on $R_{\rm SO}$, 

\noindent -- 
in model~M3, 
a higher AGN gas density of $\rho_{\rm AGN}=10^{-9}~{\rm g~cm^{-3}}$, 

\noindent -- 
in model~M4, 
a less inclined orbit with 
$\theta_*=85^\circ$, 

\noindent -- 
in model~M5, 
a thinner disk of 
$H_{\rm AGN}=5\times 10^{11}~{\rm cm}$, which is still several times thicker than that in the standard $\alpha$ disk model. This is justified as the heating rate by the stellar collisions ($\sim 10^{42}~{\rm erg/s}$ on average from the total radiated bolometric energy of $3\times 10^{46}~{\rm erg}$ divided by the orbital period in the fiducial model)
is higher than the cooling rate of the 
$\alpha$ disk model at the position of the star. 

\noindent -- 
in model~M6, 
a shorter orbital period of 
$P_{\rm orb}=5~{\rm hr}$,

\noindent -- 
in model~M7, 
a higher SMBH mass of 
$M_{\rm SMBH}=10^7~\Msun$, 

\noindent -- 
in model~M8, 
a high orbital eccentricity of 
$e=0.7$,

\noindent -- 
in model~M9, 
a shallower vertical density profile with $n=3$, 

\noindent -- 
in models~M10 and M11, 
highly inclined cases of $\theta_*=50^\circ$ and $\theta_*=0^\circ$, respectively, 

\noindent -- and finally, 
in model~M12, 
we compare our results specifically to the properties of eRO-QPE1. 
We set  
the large bow shock size to 
$R_{\rm SO}=6\times 10^{11}~{\rm cm}$ assuming a star with a mass of $\sim 30~\Msun$ \citep{Torres10}
and adopt a low-inclination orbit with $\theta_*=87^\circ$. 
The lightcurves in the X-ray bands for 
models~M1--M10 
are presented in Fig.~\ref{fig:light_curves}.

In model~M2, due to the smaller $R_{\rm SO}$, 
the amount of gas undergoing shocks with high $\theta_{\rm rel-bs||}$
is lower than in the fiducial model. As a result, the X-ray luminosity in this model is lower (orange line in Fig.~\ref{fig:light_curves}a) by a factor of $\sim 9$ compared to the fiducial model.

In the higher AGN gas density model (M3), 
the maximum X-ray luminosity is higher (blue line in panel~a) because the breakout luminosity is proportional to the gas density. 
On the other hand, the duration is shorter because the breakout timescale in the planar phase is inversely proportional to the gas density and the luminosity decreases with time as $\propto (t/t_{{\rm BO},i,j})^{-4/3}$  (Eq.~\ref{eq:li_t}).

In the less inclined model (M4), 
the total energy of emission and the peak of $L_{\rm x}$ are both high (red line in panel~a) because 
the duration for shocks to emerge is long and the relative velocity between the AGN disk gas and the star is high.

In the thinner disk model (M5), 
the shocks break out of the disk earlier due to the shorter distance from the disk midplane to the AGN surface. 
Also, the duration of the X-ray emission is 
slightly 
shorter than in the fiducial model (cyan line in Fig.~\ref{fig:light_curves}a), 
since $t_{\rm sph}$ is shorter.

In model~M6, the high orbital velocity of the star increases both the breakout luminosity and the temperature of shocked gas. 
The same results are also seen for the high $M_{\rm SMBH}$ and high $e$ cases (models~M7 and M8, blue and red lines in panel~b), in which the orbital velocity is higher at the time of disk-crossing. 
For the models with the orbital periods of $5$ (model~M6), $10$, $20$ (model~M1), $40$, and $80~{\rm hr}$, in which the relative velocities between the star and unshocked AGN gas are 
$0.24$, $0.19$, 
$0.15$, $0.095$, and $0.075~c$, respectively, 
the peak X-ray luminosities are 
$2.4\times 10^{42}$, 
$2.0\times 10^{42}$, 
$1.6\times 10^{42}$, 
$8.1\times 10^{41}$, 
and $1.8\times 10^{41}~{\rm erg/s}$. 
The peak X-ray luminosities for $P_{\rm orb}=5$, $10$, and $20~{\rm hr}$ are similar, since the maximum peak frequency of the radiation is at around the X-ray band ($0.4$--$3~{\rm keV}$). 
On the other hand, for $P_{\rm orb}\geq20~{\rm hr}$, the peak X-ray luminosity significantly decreases as $P_{\rm orb}$ increases since 
the X-ray band is shifted into the Wien tail of the black-body like spectra and the temperature of the shocked gas is lower at a lower relative velocity. 
Thus, in the fiducial settings, 
the period of QPEs needs to be shorter than $20~{\rm hr}$ ($P_{\rm QPE}=\frac{1}{2}P_{\rm orb}$) 
to achieve $L_X\gtrsim $
several $\times 10^{41}~{\rm erg/s}$. 
Also, 
the peak of the multi-color black-body emission falls in the X-ray band and the X-ray luminosity is largest 
for 
$P_{\rm QPE}\lesssim 10~{\rm hr}$. 

In these models (M6--M8), multiple components can be found in the X-ray light-curve.
One is contributed by breakout emission in the early planar phase, which is commonly seen in the other models. 
Additionally, the emission in the spherical phase contributes a 
second 
component, 
which is relatively bright because the total luminosity and temperature of radiation at the beginning of the spherical phase are high. 
Such features may have been observed in eRO-QPE1 as discussed in $\S~\ref{sec:features_qpe}$.

In model~M9, 
the properties of the emission are similar to those in the fiducial model (dashed cyan line in panel~b). 
This is because the emission is dominated by the planar phase, and the properties of the emission in the planar phase are not affected the density profile, $n$ (e.g.~Eq.~\ref{eq:li_t}).

In models~M10 and M11, 
the X-ray luminosity is significantly lower (Table~\ref{table:properties_events}). 
This is because the 
lower relative velocity of the star and the AGN disk, 
and because the time for the star to exit the disk is short. 
Hence, the star needs to orbit in the retrograde direction with respect to 
the AGN disk to explain the  high luminosities of QPEs.

In model~M12, 
the high-X-ray luminosity and the long duration as found in eRO-QPE1 are roughly reproduced due to the large values of $R_{\rm SO}$ and $\theta_*$. 
More precise comparisons to eRO-QPE1 would require a detailed modelling of the shock dynamics, which is beyond the scope of this paper. 

Overall, we find that 
the duration of the X-ray emission is significantly influenced by 
$H_{\rm AGN}$ 
(model~M5), 
and $\theta_*$ 
(model~M4), 
and the X-ray luminosity by $R_*$ 
(model~M2) 
and $\theta_*$. 
To produce the luminosity and the duration observed in QPEs, the star needs to move in the
retrograde
direction relative to the AGN gas 
at relatively low inclinations 
($\lesssim40^\circ$) 
and be massive ($\gtrsim 10~{\Msun}$).

\section{Discussion}

\subsection{Formation process}

\label{sec:formation}

Here we discuss the possible origin of the star on a low-inclination orbit relative to the 
accretion 
disk. 
Pathways to form such systems have been proposed by \citet{Lu2022}. 
They considered that a tight orbit 
with a semi-major axis of $\sim100~{\rm AU}$ and a pericentre distance of $\sim~{\rm AU}$ can be formed by the Hills mechanism.
In this picture, a stellar binary is disrupted by the tidal field of the SMBH, resulting in formation of a star tightly bound to the SMBH and a hyper-velocity star. 
Subsequently, GW radiation circularized the stellar orbit and reduces its semi-major axis to $\sim~{\rm AU}$. 
During the GW in-spiral, if scatterings by other stars are efficient, the orbital eccentricity tends to be enhanced and the condition for (partial) tidal disruption \footnote{If part of the stellar envelope lost at pericentre is consumed within an orbital period, it is observed as a partial tidal disruption event. }
with a large semi-major axis is satisfied. 
Since scattering by stars are expected to be inefficient around less massive SMBHs, QPEs are favoured to be associated with less massive SMBHs in this scenario \citep{Lu2022}, which is consistent with observations of QPEs. 
In this process, 
the star may avoid complete disruption, 
and the remnant may then form a tight orbit 
and the stripped envelope 
may form an accretion disk, which can produce QPEs. 
If this model is correct, it demonstrates that stars presumably often survive tidal disruption events (especially for low-mass SMBHs), 
since two QPE events have been observed to follow tidal disruption events \citep{Miniutti2019,Chakraborty2021}. 
Note that the lifetime of QPEs in these scenarios is limited by the viscous timescale ($t_{\rm vis}$) of the accretion disk, which is typically longer than the observed duration of QPEs as 
\begin{eqnarray}
\label{eq:t_vis}
t_{\rm vis} \sim 2\times 10^2~{\rm yr}
\left(\frac{t_{\rm QPE}}{10~{\rm hr}}\right)
\left(\frac{a/H_{\rm AGN}}{100}\right)^{2}
\left(\frac{\alpha}{0.1}\right)^{-1}.
\end{eqnarray}

It is also possible that a star has become tightly bound to the SMBH through the Hills mechanism or other relaxation processes, while an accretion disk forms by unrelated processes such as gas accretion or tidal disruption of another star. 
In those cases, the accretion disk need not be necessarily reoriented as discussed below.

The model proposed by \citet{Lu2022} requires moderate orbital eccentricity in the phases producing QPEs. On the other hand, our model allows 
all the way down to zero 
eccentricity, 
but a low-inclination and retrograde orbit with respect to the 
accretion 
disk are required to produce bright emission as seen above. 
The processes to produce less inclined orbits are discussed in the next section.

\subsection{Comparisons to observed events}

\label{sec:features_qpe}

We next discuss the consistency (or lack thereof) between our model and several observational features. 

First, QPEs have been observed with a range of X-ray energies and flare durations.
Fig.~\ref{fig:l_t} shows the distribution of $L_X$ and the duration for the observed events (filled circles) and 
models~M1--M8, M10, and M12 (crosses). 
The X-ray luminosity is significantly influenced by $\theta_*$, $\rho_{\rm AGN}$, and $R_{\rm SO}$, 
and the duration of emission is influenced by $\theta_*$, $\rho_{\rm AGN}$, and $H_{\rm AGN}$.

We expect that $\theta_*$ is distributed over a wide range. 
This is because 
the stellar-orbit at $\sim 100~r_{\rm g}$ is in the plane in which the Hills mechanism occurred, which is presumably randomly oriented.
On the other hand, 
the Bardeen-Petterson effect aligns the 
accretion 
disk with the spin direction of the SMBH 
within the radii \citep{Natarajan1998}
\begin{eqnarray}
\label{eq:r_w}
r_{\rm w} \sim 230~r_{\rm g}
\left(\frac{a_{\rm SMBH}}{0.9}\right)^{2/3}
\left(\frac{a/H_{\rm AGN}}{100}\right)^{4/3}
\left(\frac{\alpha}{0.1}\right)^{2/3},
\end{eqnarray}
where $a_{\rm SMBH}$ is the dimensionless SMBH spin. 
In the collision scenario, $r_{\rm w}$ can limit $P_{\rm QPE}$, and $r_{\rm w}$ in \eqref{eq:r_w} is roughly consistent with 
$a\sim 70$--$360~r_{\rm g}$ calculated from $P_{\rm QPE}$ and $M_{\rm SMBH}$ for the observed events (Table~\ref{table:properties_events}). 
Although recent GRMHD simulations derived smaller radii for alignment by the Bardeen-Petterson effect \citep{Liska2019,Liska2021}, these values may be underestimated due to the strong magnetic field amplified due to the initially thick disk. 
Our model requires 
$\theta_*\gtrsim 50^\circ$ 
to produce QPE flares as bright as $\gtrsim$ a few $\times 10^{41}~{\rm erg/s}$ 
(with the fiducial parameters otherwise). 
This reduces the rate of events by about 
a factor of several 
compared to the formation of these systems with all relative orientations. 

The finding that the luminous flare in eRO-QPE1 has longer duration \citep{Chakraborty2021} is consistent with this event having a lower inclination (larger $\theta_*$) compared to the other events. 
The range of the maximum value of $L_{\rm X}$ and the duration in the observed events can also be attributed to variations of the model parameters, $\theta_*$, $R_{\rm SO}$, $\rho_{\rm AGN}$, and $H_{\rm AGN}$ (Fig.~\ref{fig:l_t}), since the values of these parameters have not been constrained by observations. 
Conversely, the QPEs provide novel constraints on these parameters.

\citet{Miniutti2022} and \citet{Chakraborty2021}, respectively, reported that the luminosity for the events GSN069 and XMMSL1 gradually decrease by an order of magnitude in $\sim 2$ and $\sim 10$ years. 
Recently, after the disappearance of QPEs in GSN069, their reappearance has been reported \citep{Miniutti2023_reQPE}. 
Prior to this reappearance, the quiescent luminosity is again enhanced, possibly due to additional partial disruption or intense mass loss from a star. 
This suggests that QPEs occur when the accretion rate is lower than some critical value. 
In our scenario, this criterion can arise as follows. 
The gas additionally lost from the star is initially aligned with the stellar orbit. 
Due to the vigorous mass loss, 
a thicker disk forms as the disk thickness is proportional to the accretion rate ($H_{\rm AGN}\propto L_{\rm AGN}/\eta$). 
For a thicker disk, the alignment radius by the Bardeen-Petterson effect can become smaller than the distance of the star from the SMBH ($r_{\rm w}\propto H_{\rm AGN}^{-4/3}$, Eq.~\ref{eq:r_w}). 
Then, at the stellar position, the accretion disk is not fully aligned with the SMBH spin, and the QPE brightness is reduced due to the larger inclination angle. 
In our model, the change in the inclination angle of the accretion disk by 
$\gtrsim 30^{\circ}$ 
can explain the inactive and active phases of GSN069, between which the luminosity of QPEs is changed by a factor of $\gtrsim 10$. 
After the accretion rate decreases below the critical value, the alignment radius exceeds the stellar distance to the SMBH, and the QPEs become brighter again.

\citet{Miniutti2022} also found that the recurrence time for GSN069 is longer when a stronger QPE is followed by a weaker QPE, and shorter vice versa.  On the other hand, \citet{Chakraborty2021} found the opposite trend for XMMSL1. 
We propose that these modulation can be explained by a low orbital eccentricity. 
When the stellar orbit is eccentric, the recurrence time between two collisions around the pericentre is shorter than that around the apocentre. 
Also, emission from shocks around a star moving toward an observer is brighter than that moving away from the observer due to Doppler beaming effects. Hence, slightly eccentric orbits possibly explain the modulation of the recurrence time between QPEs found for GSN069 and XMMSL1.

The luminosity in the quiescent phases of GSN069 also oscillates roughly at the orbital period in GSN069 \citep{Miniutti2022}. 
We propose that this can be caused by emission from shocks emerging due to the dissipation of the stellar envelope unbound due to collision with the 
accretion 
disk, as proposed by \citet{Lu2022}. The diffusion timescale of the shocks within the 
accretion 
disk is about $\sim 10~{\rm hr}$, being consistent with the observed period of the modulation in the quiescent phases. 

Finally, for eRO-QPE1, highly complex evolutions have been found both for the luminosity and the temperature in some phases \citep{Arcodia2022}. 
We propose that three distinct peaks can arise in the light-curve,  
from emission with high temperature in planar phases, as discussed above for models M6--M8. Other peaks can be contributed by emission from shocks caused by the unbound stellar envelope as presented in \citet{Lu2022}. To our knowledge, this is the only proposed explanation for the multiple peaks seen in this event.

\subsection{Caveats}
\label{sec:caveats}

We have ignored possible absorption and scattering by gas far from the region producing the emission.  If the angle between the shock surface and the AGN disk is small, the shocked gas does not go round and intersect the preceding shocks \citep{Linial2019}.  In this case, the optical depth of the shells producing the observed photons is not influenced by the shock on the far sides of the star (the receding shock). Also, the emission from a shock is not obstructed by a preceding shock, as the preceding shock is not expanding fast enough to cover the luminous shell of the receding shock. 
Even if the shocked gas goes round the preceding shocks in later phases, since the density of the wrapping gas is much lower than the AGN density, 
we presume that the optical depth of the breakout emission from the preceding shocks is not significantly enhanced by the possible intersection. 
Thus, our assumption that emission from shocked gas is not obscured by shocked gas elsewhere in the disk appears at least reasonable. 

Additionally we have ignored the acceleration of shocked gas in the direction of the motion of the star \citep{Matzner2013,Irwin2021}.  
To our knowledge, the impact of this lateral acceleration on the structure of the emerging, expanding cloud, and the corresponding 
light-curve of the breakout emission, have not been investigated in literature. 
In near spherically symmetric geometry, 
the shocked gas that expanded earlier can 
obscure the breakout emission from gas that shocked later, reducing the overall luminosity. 
On the other hand, in the planar geometry considered in our model, 
such obscuration may be avoided since the shocked gas may not cover the later breakout emission \citep{Linial2019}. 
Complex light curves in eRO-QPE1 may also be related to absorption and re-emission by expanded shocked gas in some phases. 
To fully check the validity of 
neglecting this acceleration,
radiative-hydrodynamical simulations are desired.

To evaluate the thermalization of the shocked gas, we have considered
photons produced by free-free emission, following \citet{Nakar2010}. 
On the other hand, in high-metallicity environments such as AGNs, 
bound-free emission dominates the free-free process. 
Thus, the thermalization process is underestimated in this study, which may somewhat affect the evolution of the temperature in the flares. 
Also, to calculate the diffusion of photons, the opacity is obtained assuming electron scattering, 
since we consider emission from inner regions of the AGN disk where gas is ionized.  However, flares in outer regions of the AGN disk 
should be different values of the opacity \citep{Thompson05,Jiang2020}. 

If the alignment of the star with the AGN disk by gaseous interactions is faster than the alignment of the disk to the SMBH spin by the Bardeen-Petterson effect, bright flares due to the stellar collisions are not expected. 
Here, we consider the evolution of the stellar orbit by the geometric drag \citep{Generozov23}. 
By assuming that the fraction of the velocity change for the star by the drag is the same as the fraction of the mass that the star sweeps within $R_{\rm SO}$ to the stellar mass, the alignment timescale in the fiducial model is estimated to be 
$\sim P_{\rm orb} m_* {\rm cos}(\theta_*)/(2 \pi R_{\rm SO}^2 H_{\rm AGN} \rho_{\rm AGN}) \sim 50~{\rm kyr}$. 
Since the alignment timescale is much longer than the mass loss timescale, the alignment of the star can be ignored in the model.

\section{Conclusions}

In this paper, we have evaluated properties of emission from flares caused by bow shocks due produced during a collision between a star and an AGN disk. We have included the emission from parts of the bow-shock that are about to emerge from the AGN disk's surface, as well as the parts which have broken out and evolved to an expanding, quasi-spherical cloud. We discussed whether this breakout emission from the bow shocks can explain the properties of the QPE events. Our main results can be summarised as follows:

\begin{enumerate}

\item
The two main features -- i.e. the duration of an $\sim$hour and the peak X-ray luminosity of $\gtrsim 10^{41}~{\rm erg/s}$ -- observed in the five QPE events can be explained by the breakout emission from the bow shocks. 

\item 
To produce luminous emission, the star needs to be 
massive ($\gtrsim 10~\Msun$) and 
on
a retrograde and low-inclination 
($\lesssim 40^\circ$) 
orbit with respect to the AGN disk. This is required in order to produce a long bow-shock with a large surface area.

\item 
Various observed features of QPEs, including
the complex luminosity evolution, 
the gradual decrease of the luminosity 
of the flares over several years, 
the evolution of the X-ray hardness ratio, 
the fluctuations in the luminosity in the quiescent phases, 
and the relatively low masses of the central SMBHs, 
can be qualitatively explained in this model. 

\end{enumerate}

As we were finalizing this paper for submission, we became aware of two similar studies posted on the arXiv,  \citep{Linial2023,Franchini+2023}. Both investigate the cooling envelope emission from expanding shocked gas formed by the collision between a secondary and an AGN disk, with \citet{Linial2023} focusing on a secondary star, and \citet{Franchini+2023} favouring a $\sim 100~{\rm M_\odot}$ BH over a star. 
In our paper, in addition to the cooling, initially optically thick expanding envelope, we also consider the emission from the bow-shock as it emerges from the surface of the AGN disk. We find that this bow-shock emission can dominate the total X-ray luminosity. 

To explain the multiple peaks in eRO-QPE1, we suggest that 
the breakout emission is required to contribute the initial flares, with secondary contributions from the cooling envelope. 
Also, in our fiducial model, shocked gas becomes optically thin before it expands by an order of magnitude. 
On the other hand, in the cooling emission model, since the expanding gas is predicted to remain optically thick until the size of the shocked gas becomes comparable to the semi-major axis of the stellar orbit, the expanding gas may absorb and influence the emission from the AGN disk, which could yield additional observational signatures. 
Note that in the cooling emission model, the thermal coupling coefficient ($\eta$, e.g. \citealt{Nakar2010}) and accordingly the X-ray luminosity may be significantly overestimated. In \citet{Linial2023}, $\eta$ is estimated after shocked gas adiabatically expands and just before photons deep inside the shocked gas diffuse out. However, $\eta$ is much lower before the shocked gas adiabatically expands and $\eta$ does not increase before photons diffuse\footnote{
This is because both the number density of photons required to establish thermal equilibrium \citep{Nakar2010} and the number density of photons (whose ratio is the definition of $\eta$) during the slow diffusion phases decrease with time with the same power-law (as $\propto t^{-3}$). 
}. 
Meanwhile, in \citet{Franchini+2023}, the luminosity may be overestimated since the photon diffusion velocity is slower than the expansion velocity of shocked gas in their fiducial model, which needs to be considered as in \citet{Linial2023}.

Nevertheless, 
our conclusions are similar to those of \citet{Linial2023} and \citet{Franchini+2023}, namely that collisions between a star and an AGN disk offers a viable explanation for observed QPE flares.

\section*{acknowledgments}

We thank Shmuel Gilbaum for reproducing our numerical results which led us to identify and correct two bugs in our code, 
and 
Brian Metzger, Itai Linial, and Yuri Levin for useful discussions. 
This work was financially supported 
by Japan Society for the Promotion of Science (JSPS) KAKENHI 
grant Number JP21J00794 (HT) and by NASA grant 80NSSC22K082 and NSF grant AST-2006176 (ZH).

\section*{Data availability}

The data underlying this article are available in the article.

\bibliographystyle{aasjournal}
\bibliography{agn_bhm}

\end{document}